\documentclass[journal,twoside]{IEEEtran}
\ifCLASSINFOpdf
\else
\fi
\usepackage{verbatim}
\usepackage{amsmath}
\usepackage{amssymb}
\usepackage[slantedGreek]{mathptmx}
\usepackage{threeparttable}
\usepackage{cite}
\usepackage{enumerate}
\usepackage{color}
\usepackage{psfrag}
\usepackage{ifpdf}
\usepackage{multirow}
\usepackage{setspace}
\usepackage{url}
\usepackage{algorithmic}
\usepackage{algorithm}
\usepackage{booktabs}
\usepackage{array}
\usepackage{amsfonts}
\usepackage{graphicx}
\usepackage[caption=false,font=footnotesize]{subfig}
\usepackage{rotating}
\usepackage{textcomp}
\usepackage{tabularx}

\newcolumntype{L}[1]{>{\raggedright\arraybackslash}p{#1}}
\newcolumntype{C}[1]{>{\centering\arraybackslash}p{#1}}
\newcolumntype{R}[1]{>{\raggedleft\arraybackslash}p{#1}}

\begin{document}
\title{DeepcomplexMRI: Exploiting deep residual network for fast parallel MR imaging with complex convolution}

\begin{comment}
\author{Michael~Shell,~\IEEEmembership{Member,~IEEE,}
        John~Doe,~\IEEEmembership{Fellow,~OSA,}
        and~Jane~Doe,~\IEEEmembership{Life~Fellow,~IEEE}% <-this % stops a space
\end{comment}
\author{Shanshan Wang,
    Huitao Cheng, Leslie Ying, Taohui Xiao,
    Ziwen Ke, Hairong Zheng* %~\IEEEmembership{Senior Member,~IEEE}
    and ~Dong~Liang*,~\IEEEmembership{Member,~IEEE}% <-this % stops a space
%\thanks{M. Shell is with the Department
%of Electrical and Computer Engineering, Georgia Institute of Technology, Atlanta,
%\thanks{Manuscript received April 19, 2005; revised December 27, 2012.}}
\thanks{This research was partly supported by the National Natural Science Foundation of China (61601450, 61871371, 81830056), Science and Technology Planning Project of Guangdong Province (2017B020227012, 2018B010109009), the Basic Research Program of Shenzhen (JCYJ20180507182400762), Youth Innovation Promotion Association Program of Chinese Academy of Sciences (2019351).}% <-this % stops a space
\thanks{Shanshan Wang, Taohui Xiao and Hairong Zheng are with Paul C. Lauterbur Research Center for Biomedical Imaging, Shenzhen Institutes of Advanced Technology, Chinese Academy of Sciences, Shenzhen 518055, China.}%
\thanks{Huitao Cheng, Ziwen Ke and Dong Liang are with Research Center for Medical AI, Shenzhen Institutes of Advanced Technology, Chinese Academy of Sciences, Shenzhen 518055, China (e-mail: dong.liang@siat.ac.cn).}% <-this % stops a space
\thanks{Huitao Cheng is with Shenzhen College of Advanced Technology, University of Chinese Academy of Sciences, Shenzhen 518055, China.}% <-this % stops a space
\thanks{Leslie Ying is with Department of Biomedical Engineering and Department of Electrical Engineering, The State University of New York, Buffalo, New York 14260, USA% <-this % stops a space
}
% <-this % stops a space
}


\markboth{1st available in 2017 DCNN-MRI, 2018 DeepMRI, 2019 DeepcomlexMRI}%
{Shanshan Wang \MakeLowercase{\textit{et al.}}: DeepcomplexMRI: Exploiting deep residual networks for fast parallel MR imaging with complex convolution}



\maketitle

\begin{abstract}
This paper proposes a multi-channel image reconstruction method, named DeepcomplexMRI, to accelerate parallel MR imaging with residual complex convolutional neural network. Different from most existing works which rely on the utilization of the coil sensitivities or prior information of predefined transforms, DeepcomplexMRI takes advantage of the availability of a large number of existing multi-channel groudtruth images and uses them as \textcolor{black}{target} data to train the deep residual convolutional neural network offline. In particular, a complex convolutional network is proposed to take into account the correlation between the real and imaginary parts of MR images. In addition, the k-space data consistency is further enforced repeatedly in between layers of the network. The evaluations on in vivo datasets show that the proposed method has the capability to recover the desired multi-channel images. Its comparison with state-of-the-art method also demonstrates that the proposed method can reconstruct the desired MR images more accurately.
\end{abstract}

\begin{IEEEkeywords}
 Deep Learning, convolutional neural network, fast MR imaging, prior knowledge, parallel imaging
\end{IEEEkeywords}

% \fi
\IEEEpeerreviewmaketitle

\section{Introduction}
\IEEEPARstart{P}{arallel} imaging has been an essential technique to accelerate MR scan. With the utilization of spatial sensitivity of multiple coils in conjunction with gradient encoding, it shortens the imaging time by reducing the amount of acquired data needed for MR image reconstruction. Typical examples include sensitivity encoding (SENSE) \cite{pruessmann1999sense}, simultaneous acquisition of spatial harmonics (SMASH) \cite{sodickson1997simultaneous}, generalized auto-calibrating partially parallel acquisitions (GRAPPA) \cite{griswold2002generalized}, iterative self-consistent parallel imaging reconstruction (SPIRiT) \cite{lustig2010spirit}, parallel imaging using eigenvector maps (referred to as ESPIRiT) \cite{uecker2014espirit} and so on \cite{pruessmann2001advances,weller2013sparsity,park2005artifact,ramani2011parallel}.

Besides the physical properties of multi-channel acquisition, many endeavors have been made to use the signal properties in MR image reconstruction. Specifically, diverse prior information has been exploited and incorporated into the reconstruction formulation as regularizors. Among them, compressed sensing utilizes image sparsity as the prior information for fast MR imaging. For example, wavelet \cite{chaari2011wavelet}, total variation \cite{block2007undersampled,osher2005iterative}, joint total variation \cite{chen2013calibrationless}, nonlocal total variation \cite{qu2014magnetic}, and dictionary learning \cite{7337391} are used to promote the sparsity of the to-be-reconstructed MR images for high accelerations. In addition to sparsity, there are also other priors under consideration, such as partial separable function\cite{liang2007spatiotemporal}, low-rank \cite{haldar2016p,kim2017loraks,shin2014calibrationless,zhou2016step,he2016accelerated}, statistics distribution regularization \cite{wu2011multivariate}, manifold fitting \cite{wachinger2012manifold,nakarmi2017kernel,poddar2016dynamic}, GS model \cite{liang2003fast} and so on \cite{samsonov2004pocsense,chang2011kernel}.

Nevertheless, most traditional parallel imaging techniques only exploited prior information either directly from the to-be-reconstructed images or with very few reference images involved. Based on the fact that enormous images have already been acquired every day with similar  anatomic information and acquisition protocols, we previously proposed, for the first time, a convolutional neural network (CNN) model \cite{wang2016accelerating,wang2016exploiting} to perform off-line training and online single-channel reconstruction for undersampled single-channel data. Meanwhile and thereafter, different deep learning approaches have been developed for fast MR imaging \cite{hammernik2018learning,knoll2019assessment,sun2016deep,Zhu2017Image,Han2017Deep,eo2018kiki,sun2018compressed,quan2018compressed,schlemper2018deep,qin2019convolutional,lee2018deep, Liu2019SANTIS}. For example, there are model-based unrolling methods \cite{hammernik2018learning,knoll2019assessment,sun2016deep,aggarwal2019modl,mardani2017recurrent}, such as VN-Net \cite{hammernik2018learning,knoll2019assessment} and ADMM-Net \cite{sun2016deep}, and end-to-end learning methods such as AUTOMAP \cite{Zhu2017Image}, U-NET \cite{Han2017Deep,lee2018deep}, and so on \cite{eo2018kiki,sun2018compressed,quan2018compressed,schlemper2018deep,qin2019convolutional,mardani2019deep}. We also extended our initial work in \cite{wang2016accelerating,wang2016exploiting} to various deep networks for different image-reconstruction applications. In particular, only Ref. \cite{hammernik2018learning,knoll2019assessment,kwon2017parallel} are dedicated to multi-channel reconstruction. Among them, \cite{kwon2017parallel} uses a multilayer perceptron (MLP), which is a fully-connected network and thereby is computationally more expensive to train than a CNN.

\begin{figure*}[!t]
    \centering
    {\includegraphics[width=0.85\textwidth]{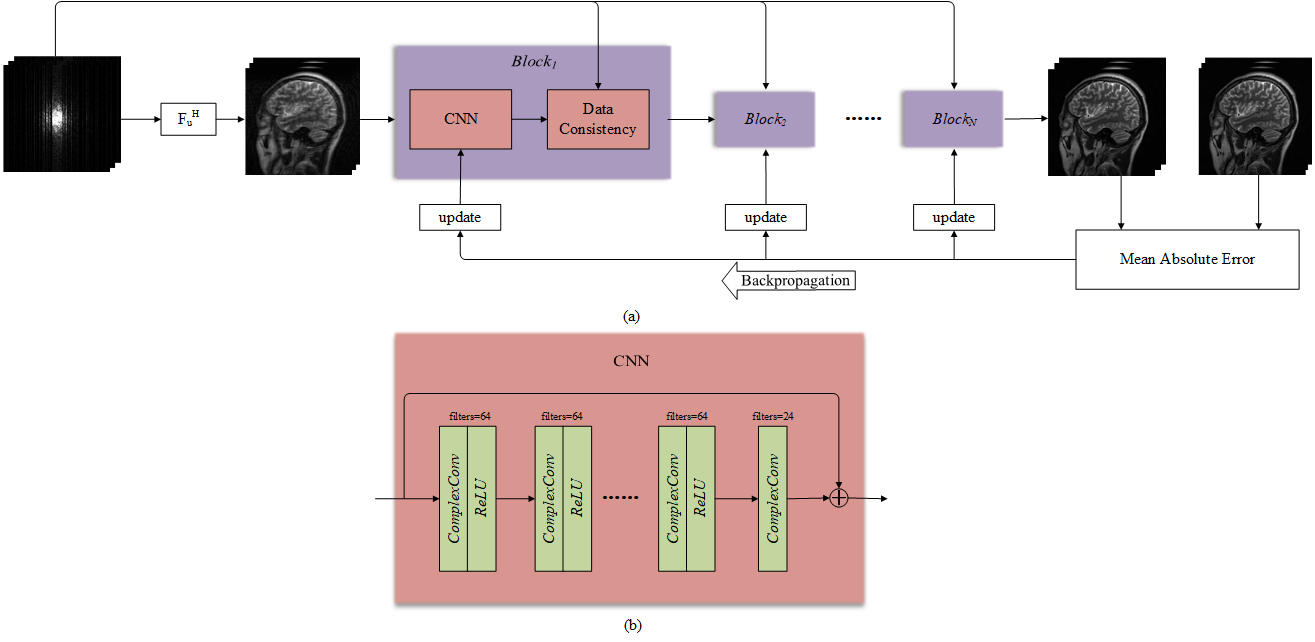}
        \label{Fig1}}
    \caption{The architecture of the proposed network for parallel imaging}
    \label{Figure1}
\end{figure*}

Motivated by the strong capability of CNN in automatic feature extraction and nonlinear correlation description for image reconstruction \cite{knoll2019assessment}, here we focus on designing a deep residual CNN for multi-channel image reconstruction (DeepcomplexMRI) \cite{wang20171d,wang2017undersampling,wang2017feasibility,wang2018investigation}. Different from the existing networks for parallel imaging, the proposed CNN provides an end-to-end network without the need for any prior information such as sparsifying transform or coil sensitivities. The network maps aliased multi-channel images to \textcolor{black}{reconstruct} multi-channel images. Such an architecture allows the image correlation across channels to be learned automatically through training. In addition, k-space data consistency is repeatedly enforced in between layers. As a result, both k-space data fidelity and image space proximity are considered in designing the network. Furthermore, complex convolutions are employed to incorporate the correlation between the real and imaginary part of MR images, which is in contrast to handling the real and imaginary parts independently for the convolution in other works \cite{wang2018complex}. Complex convolution has been used in image classification and audio-related task \cite{trabelsi2017deep} \textcolor{black}{and there has been some preliminary works investigating complex networks for fast MR imaging \cite{hui1995MRI,Virtue2017Better,Dedmari2019Complex,Cole2018Complex-valued,wang2018complex}, but its effectiveness in parallel imaging is worth investigation. This is especially important for parallel imaging because the correlations across channels are represented in both the magnitude and phase variation}. We investigate the performance of the proposed method with respect to several typical 1D and 2D undersampling patterns. And it has been tested on a series of in vivo datasets and compared to the classical parallel image reconstruction methods such as SPIRiT, and the compressed sensing based parallel imaging method L1-SPIRiT. \textcolor{black}{Specifically, our contribution can be summarized as follows: 1). We proposed an end-to-end parallel imaging reconstruction framework for MR reconstruction with convolutional neural networks exploring the multi-channel correlations. The proposed framework doesn¡¯t need any calculation of the sensitivity information to resolve the aliasing and correlations among the channels.
2). Both real-valued and complex-valued versions of our proposed framework have been investigated for parallel imaging. Complex-valued convolution network could achieve comparable and even superior performance than real-valued networks with almost only half of the real-valued network size. Our code has been released to the public.
3). The method has been compared with the classical parallel imaging methods and the supplemented deep learning based method Variational network (VN) parallel imaging work. Encouraging performances have been achieved by our proposed framework.
}

\section{Methods}

\subsection{Review of Convolutional Neural Network}

For a complete illustration of the proposed method, we firstly provide a brief review of the convolutional neural network (CNN). An L-layer CNN $y=C(x,\theta)$ can be described as follows
\begin{equation}
\label{eq_1}
\left\{
\begin{aligned}
C_{0} & = X \\
C_{l} & = \sigma_{l}(\Omega_{l} * C_{l-1} + b_{l}) & l= 1,2,...,L-1\\
C_{L} & = \Omega_{L} * C_{L-1} + b_{L}
\end{aligned}
\right.
\end{equation}
where $\Omega_{l}$ denotes the convolution operator of size $FW_{l} \times FH_{l} \times K_{l-1} \times K_{l}$ and $b_{l}$ is the $K_{l}$ dimensional bias with its element associated with a filter. The CNN output y is $C_{L}$, the output of the final layer. Here, $K_{l-1}$ is the number of the feature maps extracted at the layer $l-1$, $FW_{l} \times FH_{l}$ means the filter size and $K_{l}$ is the number of filters at layer $l$, and $\sigma_{l}$ means the nonlinear mapping operator. Eq. \ref{eq_1} can be regarded as the forward pass of the CNN training, where the convolution operator is used to extract the features and $\sigma_{l}$ calculates the nonlinear activation.

Besides the forward pass, the backward propagation updates the network parameters by calculating the backward gradients. Specifically, given the training pairs $(x,y)$, the backward propagation is to update $\theta = {(\Omega_{1}, b_{1}),...,(\Omega_{l}, b_{l}),...,(\Omega_{L}, b_{L})}$, which minimizes a cost function
\begin{equation}
\label{eq_2}
\widehat{\theta} = \mathop{\arg\min}_{\theta} \\ J(x,y)
\end{equation}
where $J(x,y)$ is the loss function determined by the specific problem. Once $\widehat{\theta}$ is obtained, it can be used for online testing tasks to predict the target $y=C(x,\widehat{\theta})$, where $x$ is now the online testing input.

\begin{figure*}[!t]
	\centering
	{\includegraphics[width=0.85\textwidth]{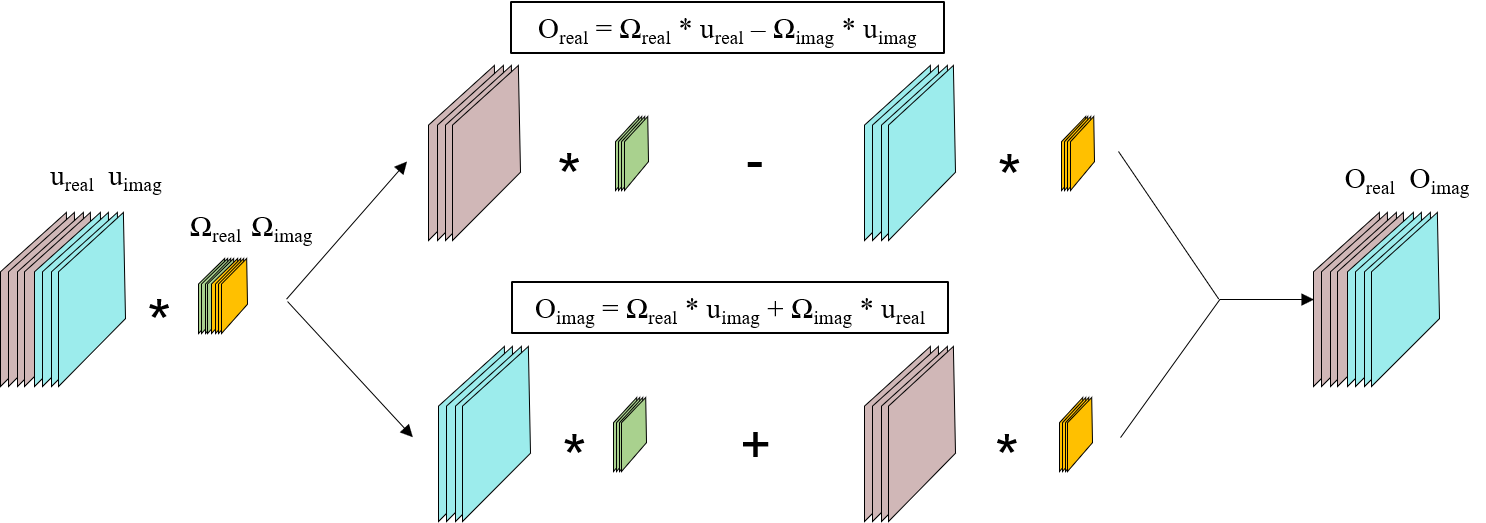} \label{Fig2}}
	\caption{The proposed complex convolution}
	\label{Figure2}
\end{figure*}

\subsection{Proposed Method}
\subsubsection{Network architecture} This work proposes a deep cascade network for parallel imaging to identify the relationship between the multi-channel undersampled MR data and the multi-channel MR image from fully sampled MR data.  Fig. \ref{Figure1} shows the architecture of the proposed network. The input of the network is the aliased multi-channel image obtained from the zero-filled undersampled k-space data, and the output is the multi-channel image reconstructed from the fully sampled k-space data. Different from the single-channel reconstruction, the input and output directly include all channels, which automatically takes into account the correlation across channels. In between the cascades of layers of the network, the acquired undersampled k-space data is used to update the layer output to enforce data consistency. It is worth noting that such a procedure is only possible when performing convolution on the full FOV images instead of patches. As shown in Fig. \ref{Figure1}, the network consists of a cascade of blocks. Each block includes a CNN unit followed by a data-consistency unit. The CNN unit is a complex convolutional network with five layers as described in Eq. \ref{eq_1}. It is worth noting that the weights of each CNN in different blocks are not shared with each other. A $3\times 3$ kernel size is adopted for each complex convolution layer. In each CNN, all the complex convolution layers have 64 feature maps except that the number of feature maps of last layer is set to the number of concatenated real and imaginary channels of the data. As for the activation function, Rectified Linear Unit (ReLU) is used after each complex convolution layer except the last layer. It is worth noting that although the number of residual blocks $n_{b}$ and the number of convolutional layers $n_{c}$ in each CNN can be adjusted, we fixed these parameters in all our experiments by setting $n_{b}=10, n_{c}=5$.

During training, the weights in each CNN are estimated using the training set by minimizing a loss function. The mean absolute error (MAE) between the ground truth and the network output was used as the training loss function, which can be described as follows
\begin{equation}
\label{eq_3}
\widehat{\theta} = \mathop{\arg\min}_{\theta} \\ J(x,y) \quad \mbox{where} \ J = \frac{1}{M}\sum_{m=1}^{M} \|C(x_{m};\theta)-y_{m}\|
\end{equation}
where $x_{m}$ denotes the $m$-th training input (multi-channel image from the zero-filled k-space data), $y_{m}$ is the corresponding groundtruth output (fully-sampled multi-channel image), and $M$ means the number of images in  the training set.

\subsubsection{Data consistency in k-space}
Besides promoting the closeness between the output image from the network and the groundtruth image, we also added a data consistency unit in each block. This unit enforces  k-space data of the output image from each block to take a value \textcolor{black}{fit} that of the acquired k-space data at the sampled k-space locations. Let $f_{l}$ denote the Fourier encoding($\mathbb{F}$) of the multi-channel image reconstructed by the learned network at layer $l$, i.e. $f_{l,j}=\mathbb{F}y_{l,j}=\mathbb{F}C_{l,j}(x;\theta)$. Let $f_{l,j}(k)$ represent the value of the k-space data of the CNN output image at the index $k$ of the $j$-th channel at layer $l$. To enforce data consistency, the value of k-space data of the CNN output image is updated by
\begin{equation}
\label{eq_4}
f_{l,j}^{rec}(k)=
\begin{cases}
f_{l,j}(k) &  k\notin S \\
\frac{f_{l,j}(k)+\lambda f_{0,j}(k)}{1+\lambda} &  k \in S
\end{cases}
\end{equation}
where $f_{0,j}$ denotes the acquired k-space data at the sampled locations, $S$ is the set of sampled locations, and $\lambda$ is a weighting parameter. The lambda was set to infinity, which means that if the k-space location is sampled, its original sample value will be used. The updated k-space data is Fourier transformed to obtain the multi-channel output image for each corresponding block, which is then fed into the next block.

\subsubsection{Complex convolution} Because of the complex nature of MR images, it is essential to properly handle complex-valued data using a deep residual convolutional neural network. The majority of the current CNNs for MR reconstruction treat a complex-valued MR image as two-channeled real-valued data, with the real and imaginary components concatenated, and perform real-valued convolutions. However, such treatment does not take the correlation between the real and imaginary components into consideration. Here, we take into account the correlation by introducing a complex convolution in each CNN layer. The complex convolution simulates complex arithmetic using real-valued arithmetic internally as explained below and thus considers the correlations between real and imaginary parts.

To perform complex convolution using real-valued entities, we convolve a complex filter matrix $\Omega = \Omega_{real} + i\Omega_{imag}$ with a complex image input vector $u = u_{real} + iu_{imag}$, where $\Omega_{real}$ and $\Omega_{imag}$ are real-valued matrices and $u_{real}$ and $u_{imag}$ are real-valued vectors . Since convolution operator is distributive, we have
\begin{equation}
\label{eq_5}
\begin{split}
\Omega * u = \ (\Omega_{real} * u_{real} - \Omega_{imag} * u_{imag}) \\+ \ i(\Omega_{real} * u_{imag} +\Omega_{imag} * u_{real})
\end{split}
\end{equation}
	
For neural network learning, proper initialization is critical in reducing the risk of vanishing gradients especially in the case when batch normalization is not adopted. A complex weight has a polar form as well as a rectangular form
\begin{equation}
\label{eq_6}
\Omega = |\Omega|e^{i\theta} = \Omega_{real} + i\Omega_{imag} = |\Omega|\cos \theta + i|\Omega| \sin \theta
\end{equation}{}
where $\theta$ and $\Omega$ are the phase and magnitude of $\Omega$, respectively.
{}
According to Chi-distribution with two degrees o{}f freedom, the magnitude of the complex weight is Rayleigh-distributed assuming that the real and imaginary components are independently and identically distributed Gaussian with equal variance and zero mean. Therefore, it is reasonable to initialize the magnitude of the complex parameter $\Omega$ using the Rayleigh distribution. We then use the uniform distribution between $-\pi$ and $\pi$ to initialize the phase of $\Omega$. With the magnitude and phase multiplied, we perform the complete initialization of the complex parameters.

\subsubsection{Final reconstruction with proposed network} Once the network is trained, we obtain a set of optimal network parameters $\widehat{\theta}$, which can be directly used for reconstructing the multi-channel image by applying the new undersampled multi-channel k-space data as the input of the proposed network and generating the output. Mathematically, the reconstruction can be represented as multi-channel
\begin{equation}
\label{eq_7}
y_{multi} = C(x_{multi};\widehat{\theta})
\end{equation}
where $x_{multi}$ is the undersampled multi-channel k-space data, $y_{multi}$ is the reconstructed multi-channel image, and $\widehat{\theta}$ denotes the well-trained weights of our proposed model. For the final single-channel reconstruction, we use an adaptive coil combination method \cite{walsh2000adaptive},  although the square root of the sum-of-squares can be used as well. The code of the proposed method is available online at \url{https://github.com/CedricChing/DeepMRI}.

\begin{figure}[!t]
	\centering
	{\includegraphics[width=0.45\textwidth]{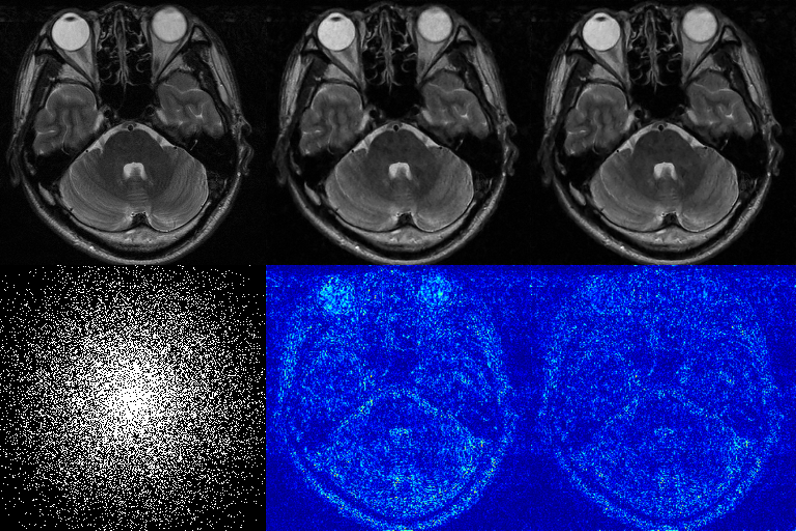} \label{Fig3}}
	\caption{Reconstruction results of proposed method with a 2D random undersampling pattern at a sampling rate of 20\%. From left to right: ground-truth (top) \& mask (bottom), reconstruction without complex convolution \textcolor{black}{(cc/0.5)} and its corresponding error map, reconstruction with complex convolution and its corresponding error map.}
	\label{Figure3}
\end{figure}

\subsection{Dataset}
\textcolor{black}{We have two datasets for evaluating our method. The training and testing datasets are all 12-channel fully sampled k-space datasets we collected from 22 volunteers with the 3T scanner (SIEMENS MANGETOM Trio Tim) and a 12-channel head coil. The images are of a great diversity including axial, sagittal, coronal images with different contrasts such as T1, T2 and PD, and of different sizes, including 256$\times$256 and 270$\times$256. For 270$\times$256 images, we cropped them into 256$\times$256 to keep all the images the same matrix size. Turbo Spin-Echo (TSE) sequence was used to acquire the dataset. For T1-weighted image, TR=928ms, TE=11ms, voxel resolution=0.9$\times$0.9$\times$0.9mm. For T2-weighted images, TR=2500ms, TE=149ms, voxel resolution=0.9$\times$0.9$\times$0.9mm. For PD-weighted images, TR=2000ms, TE=13ms, resolution=1.0$\times$1.1$\times$1.1mm. Informed consents were obtained from the imaging subject in compliance with the Institutional Review Board policy. In addition to our dataset, we also evaluated our method on the coronal proton-density (PD) data subset with the sequence named ¡°Coronal Spin Density Weighted without Fat Suppression¡± for comparison with VN method. The data were obtained from a total of 20 patients. Details could be obtained from its website https://fastmri.med.nyu.edu. 14 data were randomly selected and used as the training set, 3 as validation set and 3 as the testing set. We normalized the multi-channel data to have a maximum magnitude value of 1. Undersampled measurements were retrospectively obtained using the pre-defined undersampling masks.} %Because deep learning depends on big training datasets, which is difficult to obtain for medical images, data augmentation including rigid transformation was adopted to improve the network performance and avoid overfitting. Specifically, for each image, we apply a rotation of an angle that was randomly drawn from $[0,2\pi)$, as well as a reflection along $x$-axis and $y$-axis. }

\begin{figure}[!t]
	\centering
	\subfloat[]{\includegraphics[width=0.85\linewidth]{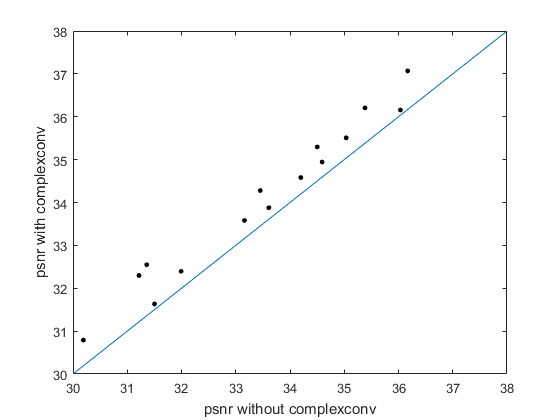} \label{Fig4-1}} \
	\subfloat[]{\includegraphics[width=0.85\linewidth]{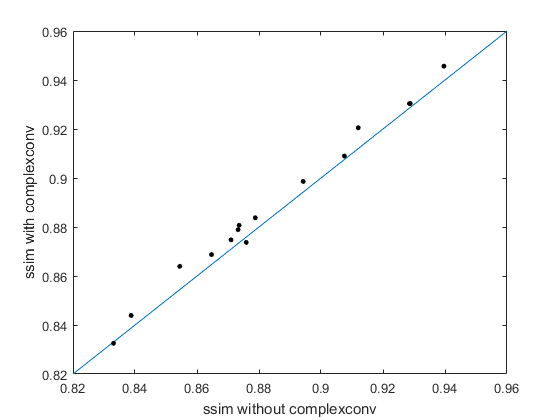} \label{Fig4-2}}
	\caption{A quantitative comparison of reconstruction results of proposed method with and without complex convolution \textcolor{black}{(cc/0.5)} in PSNR and SSIM.}
	\label{Figure4}
\end{figure}

\begin{figure*}[!t]
	\centering
	{\includegraphics[width=0.85\textwidth]{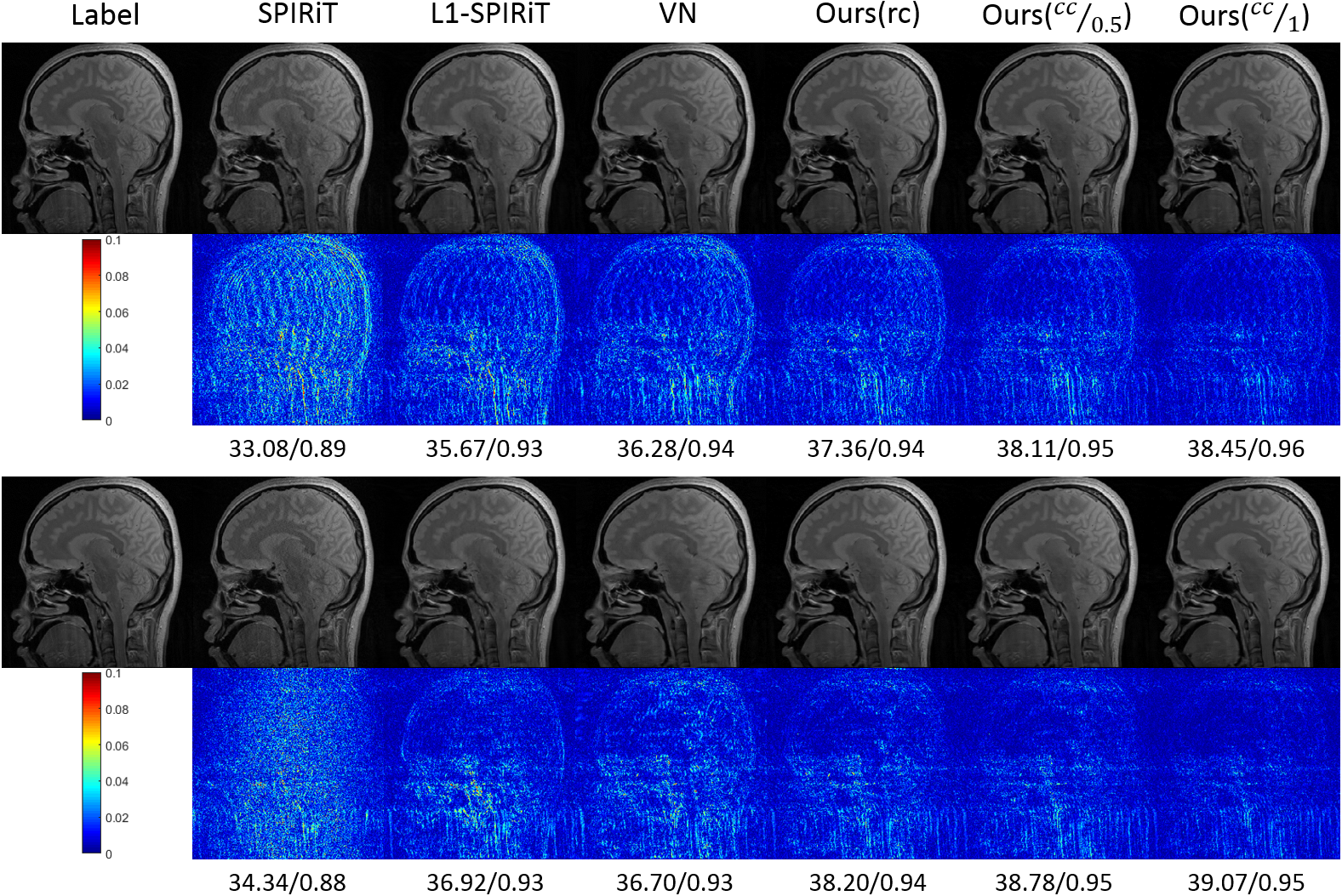} \label{Fig5}}
	\caption{\textcolor{black}{The comparison of SPIRiT, L1-SPIRiT, VN and the proposed method with real convolution (rc) and with complex convolution half parameter (cc/0.5) and with the same parameter (cc/1) under 1D random (top two lines) and uniform (bottom two lines) sampling at a sampling rate of $33\%$. The number of autocalibration line is 24 for random sampling, and 28 for uniform sampling. PSNR and SSIM values are given under the results.}}
	\label{Figure5}
\end{figure*}

\subsection{Network Configuration and Implementation}
We used a minibatch size of 4 and 40 epochs for the training of the proposed network. With respect to the optimizer, Adam method was used with $\beta_{1}=0.9, \beta_{2}=0.99$ and initial learning rate $=$ 0.0001. The training process was implemented on an Ubuntu 14.04 LTS (64-bit) operating system equipped with Graphics Quadro K40c in the open framework Tensorflow. The testing process was implemented on an Ubuntu 14.04 LTS (64-bit) operating system equipped with 128GB RAM and Intel Xeon(R) CPU E5-2660 v3 running Matlab 2015b. The testing process was also implemented with the Tensorflow GPU version. \textcolor{black}{The training time of the network is about 46 hours (with 10 blocks, 5 layers in each block, batch size 4, 40 epochs, Nvidia K40c GPU).}

\subsection{Undersampling Masks}
Four different types of undersampling patterns were tested, including 1D variable density random, 1D uniform, 2D Poisson disc, and 2D random. For 2D Poisson disc mask, 5x and 10x acceleration were simulated by retaining 20\% and 10\% raw k-space data. For both 1D random and uniform masks, 3$\times$ and 4$\times$ acceleration were simulated. For the 2D random mask, a 25\% sampling rate was simulated. It should be stated that different models were trained for each sampling pattern. Sampling pattern was fixed during training for a specific task.

\begin{figure*}[!t]
	\centering
	{\includegraphics[width=0.85\textwidth]{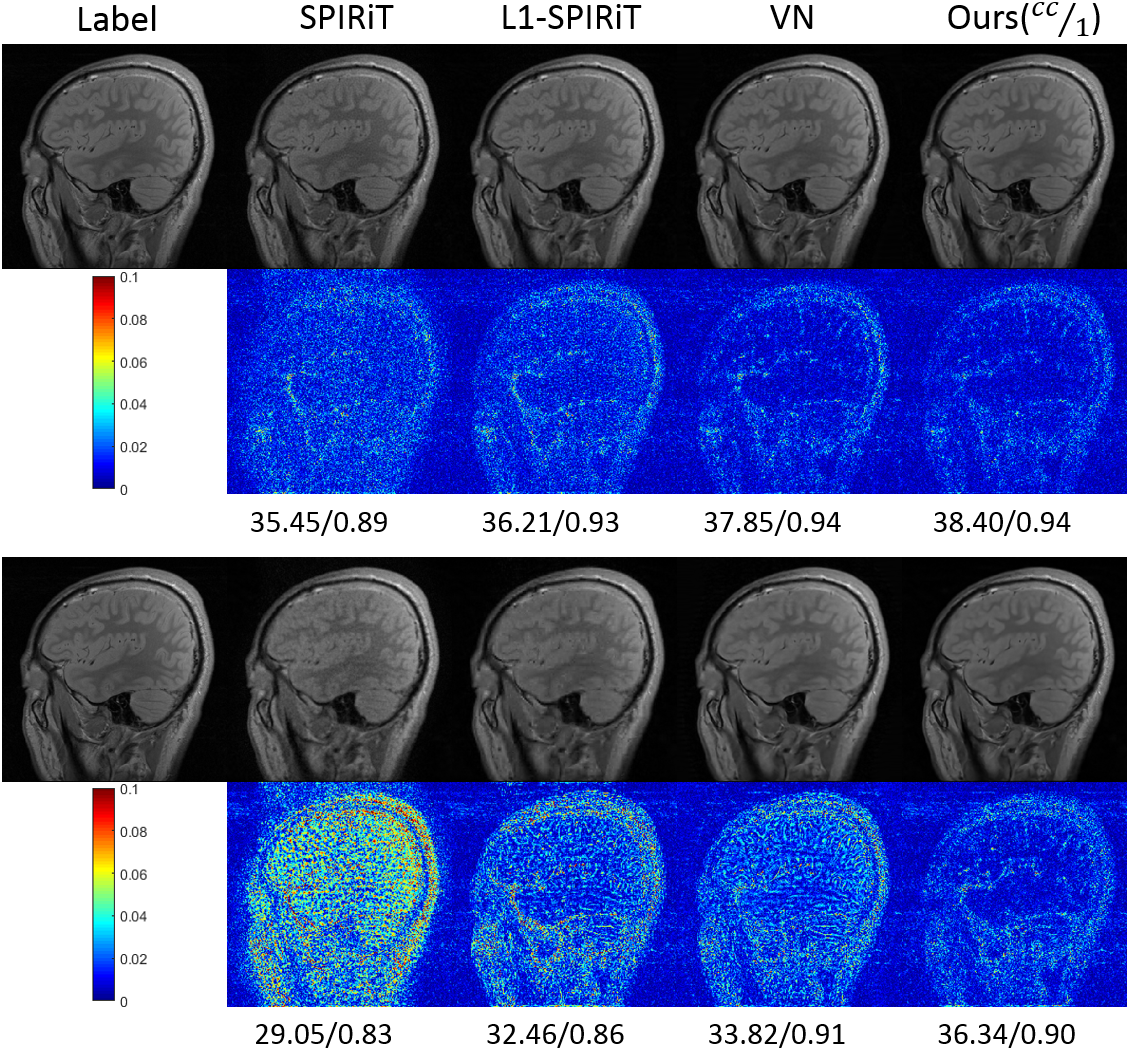} \label{Fig6}}
	\caption{\textcolor{black}{The comparison of SPIRiT, L1-SPIRiT, VN and the proposed method with complex convolutions (cc/1) with 2D poisson sampling masks at a sampling rate of $20\%$ and $10\%$. The calibration size in both masks is 30 $\times$30. PSNR and SSIM values are given under the visual results.}}
	\label{Figure6}
\end{figure*}
\begin{figure*}[!t]
	\centering
	{\includegraphics[width=0.85\textwidth]{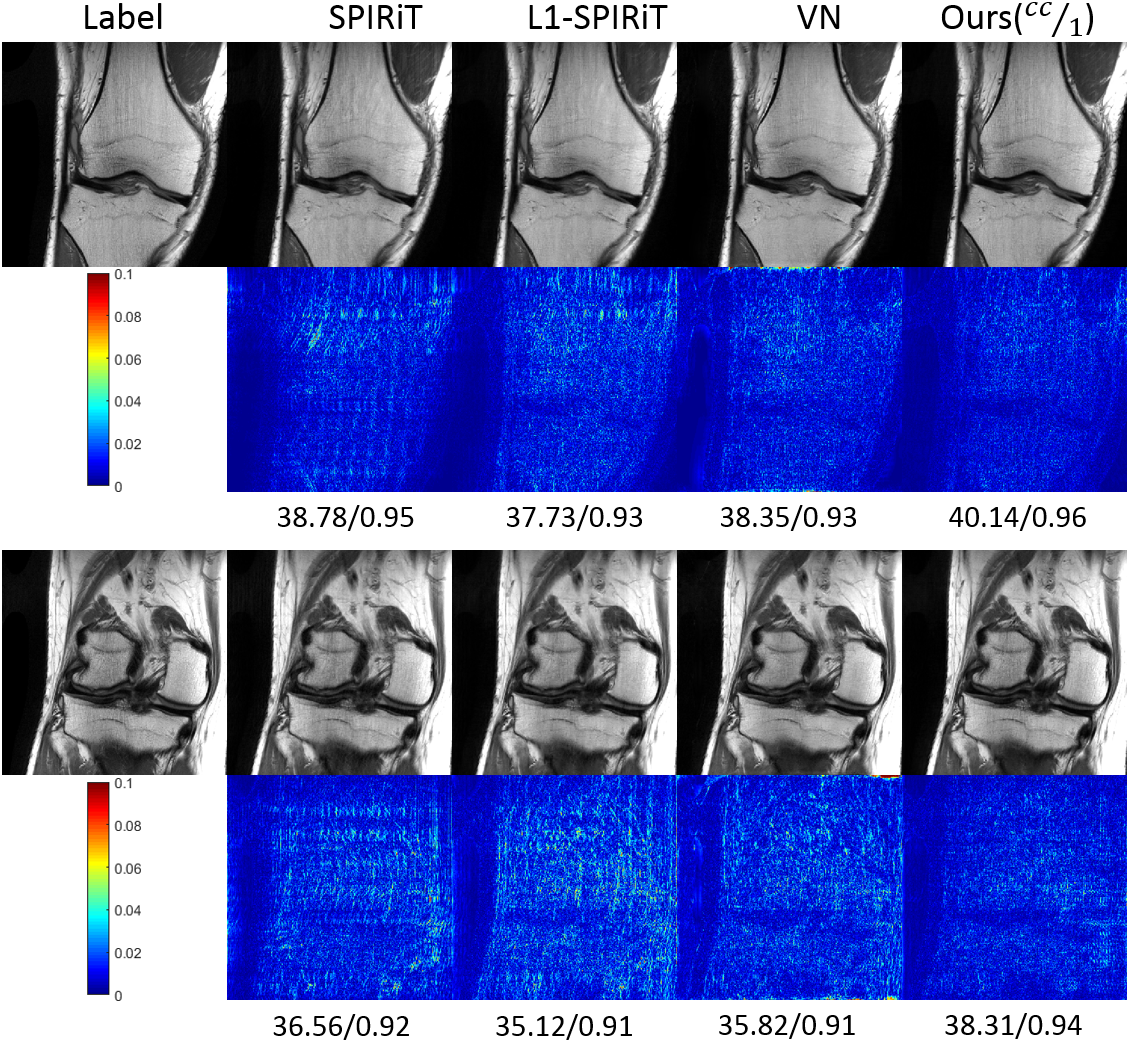} \label{Fig6R}}
	\caption{\textcolor{black}{The comparison of SPIRiT, L1-SPIRiT, VN and the proposed method with complex convolutions (cc/1) with 1D random sampling masks at a sampling rate of 33\%. PSNR and SSIM values are given under the visual results.}}
	\label{Figure6R}
\end{figure*}
%
%\begin{figure}[!t]
%	\centering
%	\subfloat[]{\includegraphics[width=0.85\linewidth]{Figure6.png} \label{Fig6}} \
%	\subfloat[]{\includegraphics[width=0.85\linewidth]{R5.png} \label{Fig6-2}}
%	\caption{(a) The comparison of SPIRiT, L1-SPIRiT, VN, proposed method with complex convolutions (cc/1) with 2D poisson sampling masks at a sampling rate of $20\%$ and $10\%$. The calibration size in both masks is 30¡Á30. PSNR and SSIM values are given under the visual results. (b) The comparison of SPIRiT, L1-SPIRiT, VN, proposed method with complex convolutions (cc/1) with 1D random sampling masks at a sampling rate of 33\%. PSNR and SSIM values are given under the visual results.}
%	\label{Figure4}
%\end{figure}
\section{Results}

\subsection{Impact of Complex Convolution}
To demonstrate the impact of complex convolution on the reconstruction, we trained two networks, between which the only difference was whether complex convolution was adopted or not.

For visual comparison, Fig. \ref{Figure3} shows the reconstruction results of proposed model with and without complex convolution. It can be seen that the reconstruction with complex convolution shows better visual quality. The error maps also indicate that network with complex convolution can achieve lower loss, which demonstrates the importance of complex convolution.

We also show the quantitative comparison in Fig. \ref{Figure4}, which also supports the use of complex convolution for proposed network based on the PSNR and SSIM measures. The result in Fig. \ref{Figure4} was the the results based on 15 complex-valued testing images with a 2D random 25\%-sampled mask. It can be observed that our proposed network with complex convolution outperforms the real convolution version of the network. Therefore, the test results in both visual and quantitative comparison demonstrated that complex convolution can help the network to obtain better image reconstruction results. To further demonstrate the effect of complex convolution on different sampling patterns, we show the results of proposed network without complex convolution for comparison in the later experiments.

\subsection{Comparison to State-of-the-art Methods}
To further evaluate the propose method, we compared our methods with \textcolor{black}{three state-of-the-art parallel imaging methods, SPIRiT, L1-SPIRiT and deep learning based parallel imaging work VN with different undersampling patterns. SPIRiT and L1-SPIRiT adopted 1D random and uniform undersampling patterns with some autocalibration lines and typical parameter settings. Specifically, the kernel size was $5\times 5$ and calibration region had 24 lines for 1D random sampling and 28 lines for 1D uniform sampling. For VN, the parameter settings were set according to their papers and code instruction.}%For SPIRiT, the iteration number was 30. As for the compressed sensing based parallel imaging technique L1-SPIRiT, Tikhonov regularization parameter was set as 0.01 and the Wavelet soft-thresholding regularization parameter was set as 0.0015.

\textcolor{black}{Fig. \ref{Figure5} shows that the images reconstructed by SPIRiT, L1-SPIRiT, VN and the proposed method at a sampling rate of 33\% with both real convolution (rc) and complex convolution (cc).  It should be noted that complex-valued architecture allows one to halve the number of total weights and so to go wider than its real counterpart for a given parameter size \cite{trabelsi2017deep}. We provide two groups results of complex-valued network. One was provided by our complex-valued network denoted as Ours (cc/0.5) which only had half size of the real-valued network denoted and a complex-valued network with the same size of real-neural networks as Ours (cc/1). It also should be noted that it is a net acceleration factor with autocalibration lines calculated for the sampling rate. It can be observed that the images reconstructed by the proposed method is closer to the ground truth image, while  the aliasing artifacts or noise is more visible in  the images reconstructed by SPIRiT andL1-SPIRiT. The corresponding error maps are also shown in Figure 5. It could be observed that there are fewer errors in our proposed method, while the other three methods show some noise and aliasing artifacts.}

To further validate our method with other sampling patterns, we also investigated our method with the 2D Poisson undersampling mask with both real and complex convolutions. We trained networks with different acceleration factors. The reconstruction results are shown in Fig. \ref{Figure6}. It demonstrates that the proposed method produced better visual quality with 2D Poisson disc masks at an sampling rate of 20\% compared to SPIRiT and L1-SPIRiT. When the acceleration factor increases to only 10\% sampling rate, the reconstruction gradually worsens. Nevertheless, the reconstruction using the proposed method is still acceptable even at 10x acceleration under 2D Poisson disc sampling. We observe that the proposed network can learn valuable prior information from the big off-line datasets, and then perform high-quality online image reconstruction from different undersampled MR data. \textcolor{black}{We also provided the comparison of SPIRiT, L1-SPIRiT, VN and the proposed method (cc/1) with 1D random sampling masks at a sampling rate of 33\% on the knee in Fig. \ref{Figure6R}. PSNR and SSIM values are given under the visual results. It can be observed that the proposed method provides encouraging performances in both quantitative and qualitative comparison.}

\subsection{SISO reconstruction vs MIMO reconstruction}
\begin{figure}[!t]
	\centering
	{\includegraphics[width=0.45\textwidth]{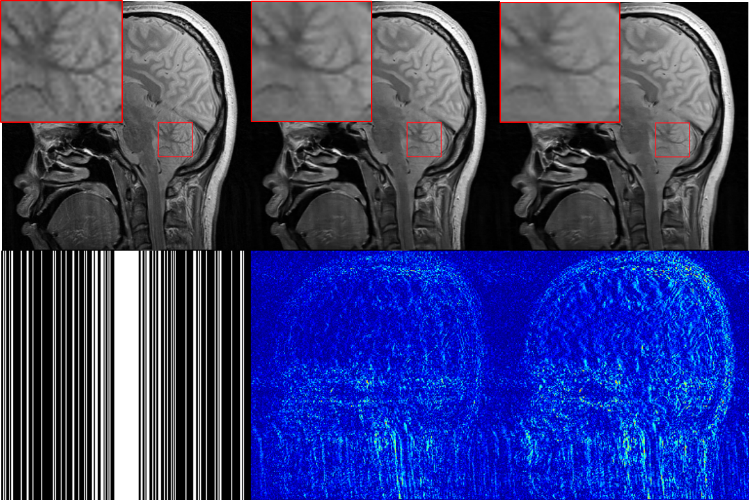} \label{Fig7}}
	\caption{A comparison of the groundtruth image (left), MIMO reconstruction (middle), and SISO reconstruction (right) with 1D random undersampling (ACS=24) at\textcolor{black}{ a sampling rate of 33\%}.}
	\label{Figure7}
\end{figure}

For parallel imaging, there are two types of networks, the combined (e.g., Walsh adaptive combination) single-channel aliased image as the input and single-channel reconstruction as the output (SISO), and multi-channel aliased image as the input and multi-channel reconstruction as the output (MIMO). Our proposed DeepcomplexMRI can be regarded as a MIMO network. Alternatively, we can construct a SISO network. Fig. \ref{Figure7} is a comparison of the results of the two methods with 1D random undersampling (ACS=24) at an acceleration factor of 3. It can be seen that the details of MIMO reconstruction are better preserved. It is apparent that the MIMO network achieves lower loss than the SISO reconstruction, demonstrating the MIMO network can actually help the network converge to a better solution. This is because the MIMO network can learn the correlations across channels while SISO uses a fixed formulation to combine multi-channel images. It suggests that training a MIMO network is preferred for parallel imaging. \textcolor{black}{On the other hand, although the performances of our proposed method are better than both classical parallel imaging and the famous deep learning based parallel imaging work, there are still some spaces left to further improve the reconstruction result compared to the groundtruth result reconstructed from the fully sampled dataset. We will devote more endeavours in the future to develop more advanced techniques for better image resolution. }

\subsection{Sensitivity to Autocalibration Lines}
\begin{figure}[!t]
	\centering
	{\includegraphics[width=0.5\textwidth]{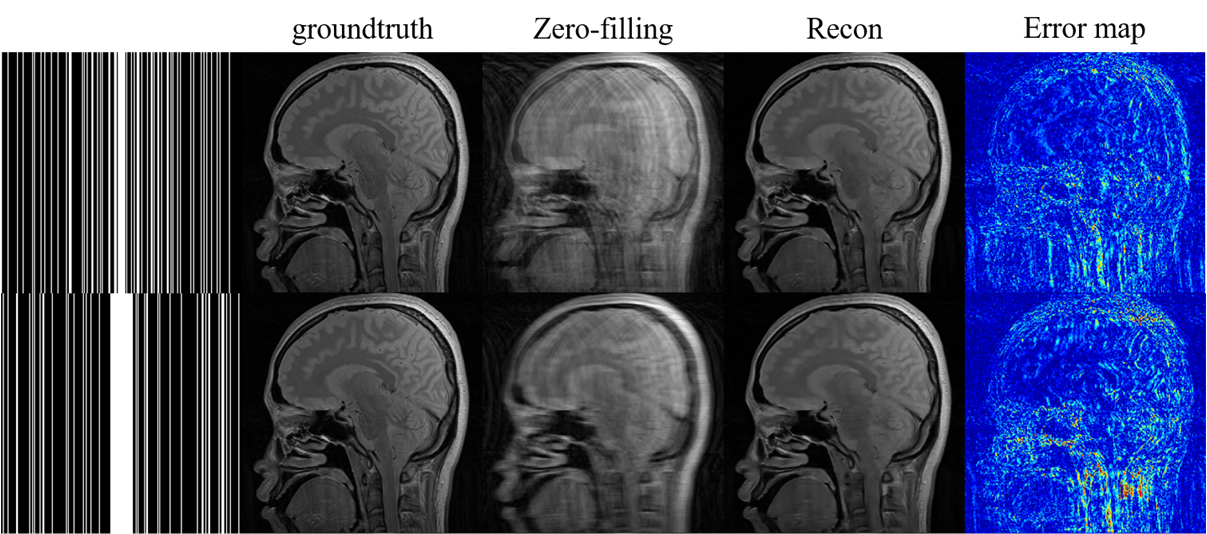} \label{Fig8}}
	\caption{From left to right: 1D random mask (AF=4, ACS=8 for first row, ACS=24 for second row), ground-truth, zero-filling image and reconstruction of the proposed method \textcolor{black}{cc/0.5}.}
	\label{Figure8}
\end{figure}

It is well known that the mask used in classical SPIRiT and L1-SPIRiT algorithms requires acquisition of autocalibration signal (ACS) lines. These classical methods are sensitive to reduced number of ACS lines, which actually prolongs the scanning time. Due to this reason, the proposed method is superior as seen in the result with 1D random mask at an acceleration factor of 3. We also performed another experiment with the same undersampling pattern at an even higher net acceleration factor (AF) of 4x but with two different types of ACS lines, namely 8 lines and 24 lines. The experimental results in Fig. \ref{Figure8} show that the proposed method can still reconstruct high-quality images with high acceleration factors even with very few ACS lines. The sensitivity of the proposed method to ACS line is relatively low.

\subsection{Convergence Property}
\begin{figure}[!t]
	\centering
	{\includegraphics[width=0.5\textwidth]{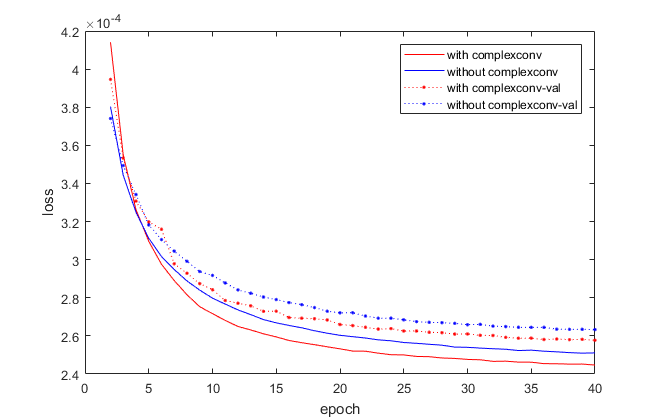} \label{Fig9}}
	\caption{The loss curves of training with and without complex convolution \textcolor{black}{(cc/0.5)}}
	\label{Figure9}
\end{figure}

The performance of complex convolution has been evaluated using the above reconstruction results in term of PSNR and SSIM. To have a better understanding about the benefit of complex convolution, we analyze the convergence property of the proposed method. Fig. \ref{Figure9} plots the loss-descending curves of the training with and without complex convolution. As can be seen from the figure, our proposed model reaches convergence after 40-epoch training. Furthermore, both the training loss and validation loss with complex convolution are lower than those without complex convolution, which further demonstrates the benefit of complex convolution. From the comparison, we claim that the improvement of the network convergence property is mainly attributed to the complex convolution. This is mainly because that complex convolution can fully leverage the magnitude and phase information based on the multi-channel data.

\section{Discussion}
In this work, we exploit deep residual network for fast parallel MR imaging with complex convolution based on our previous research. A complex residual network is designed to train multi-channel parallel MR data. Residual connections can avoid the vanishing gradient problems and complex convolution can improve convergence ability of network. By comparing the experimental results in Figs. 3 and 4, we conclude that the proposed deep residual learning network based on complex convolution can achieve better reconstruction quality.

As for the inference time, there are two different scenarios to predict a multi-channel reconstructed image using the multi-channel undersampled input. One is performing the forward process of the model based on Tensorflow with GPU acceleration, and the other is calculating the forward process on MATLAB platform with CPU. When the first way is adopted, the inference time of DeepcomplexMRI using the same setting in our experiments is about 1.5 seconds, which is much faster than the conventional methods owing to the advantage of the computational graph and GPU acceleration. If the second way is used, the inference time is about 100 seconds. Therefore, it is recommended that testing process is implemented on computational graph-based Tensorflow GPU platform.

\section{Conclusion}
This work designs a deep residual network for parallel MR imaging. The network is trained using a huge set of existing high-quality fully-sampled multi-channel data, and then serves as a predictor to reconstruct the image from the undersampled data. The experimental results have shown that the proposed method can produce images with less noise and artifacts than the state-of-the-art methods at the same acceleration factor.

\ifCLASSOPTIONcaptionsoff
  \newpage
\fi



\bibliographystyle{IEEEtran}
\bibliography{IEEEabrv,LINDBERG}
\end{document}